\documentclass[a4paper,bibtotoc]{scrartcl}

\usepackage{graphicx}
\usepackage{amsmath}
\usepackage{amssymb}
\usepackage{amsthm}
\usepackage{setspace}
\usepackage{longtable}
\usepackage[colorlinks=true,%
            linkcolor=blue,%
            citecolor=blue,%
            urlcolor=blue]{hyperref}           % for hyperlinks

\usepackage{fancyhdr}           % For LaTeX2.09 you should specify {fancyhdr}
\usepackage{ifthen}

\newcommand\settitle[2][]{%
 \title{#2}
 \ifthenelse{\equal{#1}{}}%
  {\fancyhead[RO]{\nouppercase #2 \qquad \thepage}}%
  {\fancyhead[RO]{\nouppercase #1 \qquad \thepage}}%
}

\newcommand\setauthors[2]{%
 \author{#2}
  {\fancyhead[LE]{\thepage \qquad \nouppercase #1}}%
}

\fancypagestyle{plain}{%
\fancyhf{}
}

\def\keywordsname{Keywords.}
\newenvironment{keywords}{%
      \list{}{\advance\topsep by-0.50cm\relax\small
      \leftmargin=1cm
      \labelwidth=1cm%\z@
      \listparindent=1cm%\z@
      \itemindent\listparindent
      \rightmargin\leftmargin}\item[\hskip\labelsep
                                    \bfseries\keywordsname]}
    {\endlist}
\newcommand{\Lev}{\operatorname{Lev}}
\newcommand{\dom}{\operatorname{dom}}
\newcommand{\op}{\operatorname{op}}
\newenvironment{proof2}{\begin{proof}\upshape}{\end{proof}}

\begin{document}
\theoremstyle{definition}
\newtheorem{theorem}{Theorem}
\newtheorem{definition}[theorem]{Definition}
\newtheorem{lemma}[theorem]{Lemma}
\newtheorem{corollary}[theorem]{Corollary}
%------------------------------------------------------------------------
% \settitle :
% the optional argument [...] should be only one line
% if the optional argument [...] exists, it will be used for the header
% otherwise the obligatory argument {...} will be used for header
% the obligatory argument is used as FULL title

\settitle[Discontinuity of Equilibria]
         {How discontinuous is Computing Nash Equilibria?}
% title for special papers: Category e.g. Summary, Discussion, Open Problems, ...
% choose ''Dagstuhl Seminar''or ''Perspectives Workshop''
%\settitle[seminar-number -- Paper Category]
%         {{\large seminar-number -- Paper Category} \\
%          Please Fill in Your own FULL Title \\
%         %\large{--- Perspectives Workshop ---}}
%          {\large --- Dagstuhl Seminar ---}}

%------------------------------------------------------------------------
% \setauthors :
% the first argument will be used for the header; it should be only one line
% the second argument is the FULL list of authors

% Warning: Do not use \and together with \usepackage{hyperref}
\setauthors{A. Pauly}
           {Arno Pauly \\ University of Cambridge Computer Laboratory\\
  Cambridge, UK\\
Arno.Pauly@cl.cam.ac.uk
           }
%------------------------------------------------------------------------
\date{}
\maketitle

\thispagestyle{plain}

%------------------------------------------------------------------------
%------------------------------------------------------------------------
\begin{abstract}
We investigate the degree of discontinuity of several solution concepts from non-cooperative game theory. While the consideration of Nash equilibria forms the core of our work, also pure and correlated equilibria are dealt with. Formally, we restrict the treatment to two player games, but results and proofs extend to the $n$-player case. As a side result, the degree of discontinuity of solving systems of linear inequalities is settled.
\end{abstract}

%------------------------------------------------------------------------
%------------------------------------------------------------------------
% old keywords command
% \keywords{Dagstuhl Seminar Proceedings, style, template}

%------------------------------------------------------------------------
%------------------------------------------------------------------------
\section{Introduction \label{intro}}
Both for applications and theoretical considerations, the algorithmic task of computing Nash equilibria from certain representations of games is of immense importance. A natural mathematical formulation of game theory uses the real numbers for payoffs and for mixed strategies, while classical models for algorithms require a restriction to countable sets. By imposing suitable restrictions and modifications to obtain countable problems, the  complexity of computing a Nash equilibrium for a normal form game was proven to be PPAD-complete (\cite{papadimitrioub}, \cite{dengd}).

Here we will use another approach: Instead of limiting the problem, we will extend the theory of computation. While the TTE-framework (\cite{weihrauchd}) is perfectly capable of formulating the task of computing Nash equilibria from normal form games, we will see that even the most trivial cases are discontinuous, and hence not computable.

To gain a deeper understanding of the problem, its degree of discontinuity will be studied. Mirroring an approach in the study of game theory using classical computational complexity, we will also examine other solution concepts such as correlated equilibria. While correlated equilibria seem to be computationally easier than Nash equilibria\footnote{In \cite{Zemel} several decision problems regarding Nash equilibria and correlated equilibria were compared, most of them turned out to be NP-hard for Nash equilibria and to be in P for correlated equilibria.}, we will prove that both concepts share a degree of discontinuity. Limitation to pure strategies yields a strictly less discontinuous problem, the classical problem can be solved by a cubic algorithm\footnote{There are, however, several interesting hardness results for finding pure equilibria in games (\cite{gottlob}, \cite{papadimitriouc}), originating in other representations or requiring additional properties.}.

\section{Preliminaries}
\subsection{Game Theory}
An $n \times m$ bi-matrix game is simply given by two $n \times m$ real valued matrices $A$ and $B$. Two players simultaneously pick an index, row player chooses an $i \in \{1, 2, \ldots, n\}$ and column player chooses an $j \in \{1, 2, \ldots, m\}$. Row player gets $A_{ij}$ as a reward, column player gets $B_{ij}$. We consider several solution concepts defined as equilibria, where no player has an incentive to change her strategy unilaterally.

\begin{definition}
A pure equilibrium for a $n \times m$ bi-matrix game $(A, B)$ is a pair $(i, j) \in \{1, \ldots, n\} \times \{1, \ldots, m\}$ satisfying $A_{ij} \geq A_{kj}$ for all $k \in \{1, \ldots, n\}$ and $B_{ij} \geq B_{il}$ for all $l \in \{1, \ldots, m\}$.
\end{definition}

As pure equilibria do not exist for all games, a more general notion is introduced. If both players can randomize independently over their actions, one is led to the definition of an $m$-mixed strategy as an $m$-dimensional real valued vector $s$ with non-negative coefficients and $\sum \limits_{j = 1}^m s_j = 1$. The set of $m$-mixed strategies will be denoted by $S^m$.

\begin{definition}
A Nash equilibrium for an $n \times m$ bi-matrix game $(A, B)$ is a pair $(\hat{x}, \hat{y}) \in S^n \times S^m$ satisfying $\hat{x}^TA\hat{y} \geq x^TA\hat{y}$ for all $x \in S^n$ and $\hat{x}^TB\hat{y} \geq \hat{x}^TBy$ for all $y \in S^m$.
\end{definition}

If $(\hat{x}, \hat{y})$ is a Nash equilibrium, again neither of the players can improve her payoff by changing her mixed strategy unilaterally. A famous result by \textsc{John Nash} (\cite{nash}) established that Nash equilibria in bi-matrix games always exist. By identifying a pure strategy with the mixed strategy that puts weight 1 on it, pure equilibria can be considered a special case of Nash equilibria. An even more general solution concept can be obtained by allowing the individual player's randomization processes to be correlated (\cite{aumann2}).

\begin{definition}
A correlated equilibrium for a $n \times m$ bi-matrix game is a real valued $n \times m$ matrix $C$ with non-negative entries and $\sum \limits_{i = 1}^n \sum \limits_{j = 1}^m C_{ij} = 1$ so that $$\sum \limits_{j = 1}^m A_{ij}C_{ij} \geq \sum \limits_{j = 1}^m A_{lj}C_{ij}$$ holds for all $i, l \in \{1, 2, \ldots, n\}$ and $$\sum \limits_{i = 1}^n B_{ij}C_{ij} \geq \sum \limits_{i = 1}^n B_{ik}C_{ij}$$ holds for all $j, k \in \{1, 2, \ldots, m\}$.
\end{definition}

Given a Nash equilibrium $(x, y)$, a correlated equilibrium can be constructed as $C_{ij} = x_iy_j$, while each correlated equilibrium of this form can be obtained from a Nash equilibrium, allowing us to consider Nash equilibria as special cases of correlated equilibria. Thus, finding a correlated equilibrium has to be easier than finding a Nash equilibrium, as we just presented a reduction.

Another way of creating an easier problem consists in a restriction of the games used. A zero-sum game is a bi-matrix game of the form $(A, -A)$.
\subsection{Representing Games}
In order to consider games as inputs to Type-2-Machines, they have to be coded into infinite sequences. The choice of the countable alphabet used is irrelevant for the theory, to simplify proofs we will use either $\{0, 1\}$ or $\mathbb{N}$, depending on the context. The degrees of discontinuity we study are those of the realizations, that is of functions turning names of instances into names of solutions. Since all occurring representations will be admissible, topological properties carry over between sets of games and sets of names for games, etc.

As games in normal form are pairs of real matrices, and (possible) equilibria pairs of real vectors (or again real matrices), one can quickly derive suitable representations by using product and coproduct representations (\cite{weihrauchd}, \cite{paulyreducibilitylattice}), starting from any representation of the real numbers.

The standard representation $\rho$ of the real numbers is chosen for various reasons; it is admissible and provides a convincing class of computable functions, in contrast to some of the alternatives (\cite{weihrauchd}, \cite{weihrauchb}). Additionally, as demonstrated in \cite{paulycca}, the representation $\rho$ is equivalent to the representation naturally arising for the results of repeated physical measurements. For defining $\rho$, we fix a bijection $\nu: \mathbb{N} \to \mathbb{Q}$ with $\nu(0) = 0$, so that all the usual operations on $\mathbb{Q}$ are computable w.r.t. $\nu$.
\begin{definition}
Let $\rho(w) = x \in \mathbb{R}$ hold for $w \in \mathbb{N}^\mathbb{N}$, if $|\nu(w(i)) - x| \leq 2^{-i}$ holds for all $i \in \mathbb{N}$.
\end{definition}

\begin{definition}
Let $w$ be a $\Gamma$-name for the bi-matrix game $(A, B)$, if \begin{enumerate}
\item $w = 0^n1^m0w_2$, when $(A, B)$ is an $n \times m$ game
\item $w_2 = \langle w^a, w^b \rangle$, where $\langle \ \rangle$ denotes the usual pairing function
\item $w^a = \langle w_{11}^a, \ldots w_{n1}^a, w_{12}^a, \ldots, w_{nm}^a \rangle$
\item $w^b = \langle w_{11}^b, \ldots w_{n1}^b, w_{12}^b, \ldots, w_{nm}^b \rangle$
\item $\rho(w_{ij}^a) = A_{ij}$
\item $\rho(w_{ij}^b) = B_{ij}$
\end{enumerate}
\end{definition}

Representations for pure, Nash and correlated equilibria can be derived in the same fashion. Detailed definitions are omitted here.

\subsection{Comparing Discontinuity}
As games can have multiple equilibria, we do not consider a function assigning an equilibrium to each game, but rather a multi-valued function. We will identify a multi-valued functions with the set of its choice functions. To compare the discontinuity of such sets, Type-2-Reducibility as studied in e.g. (\cite{weihrauchc}, \cite{mylatz}, \cite{stein}, \cite{hertling}, \cite{brattka}, \cite{paulymaster}, \cite{paulyreducibilitylattice}) is used, as well as the Level of a function (or a set of functions), introduced in \cite{hertling}.

We use the following definition of Type-2-Reducibility:

\begin{definition}
Let $A, B$ be multi-valued functions. Then $A \leq_2 B$ holds, iff there are continuous partial functions $F$, $G$ with $w \mapsto F(w, g(G(w))) \in A$ for each $g \in B$.
\end{definition}

As demonstrated in \cite{paulyreducibilitylattice} (for suprema) and \cite{brattka2} (for infima), $\leq_2$ induces a completely distributive complete lattice. We use $\lceil P_n \rceil_{n \in \mathbb{N}}$ to denote the supremum of a countable family $(P_n)_{n \in \mathbb{N}}$. This allows to consider the degree of discontinuity of finding equilibria in any game as the supremum of the degrees of discontinuity of finding equilibria in games with fixed size.

As the Level will play only a minor role in our considerations, we refer to \cite{paulyreducibilitylattice} for definitions.

\section{Single Player Games and Pure Equilibria}
From the perspective of game theory, single player games are trivial: The acting player chooses whatever action is best for her. As a discrete computation problem, this amounts to finding a maximum in a list of integers, a task that can be solved in linear time or logarithmic space. As the problem posed over the reals is discontinuous, we will study the problems $\textsc{1Pure}_n$ and $\textsc{1Pure}$ of finding pure equilibria in single player games with $n$ actions and without fixed game sizes. It shall be noted that single player games can be identified with $n \times 1$ bi-matrix games, justifying their inclusion.

As every $n \times 1$ bi-matrix game has a pure equilibrium, and $C_{ij} > 0$ can only hold in a correlated equilibrium $C$, if the entry $A_{i1}$ is maximal in $A$ (and thus $(i, 1)$ is a pure equilibrium), finding pure, Nash and correlated equilibria is equivalent for single player games, so the restriction to pure equilibria does not invoke any loss of generality.

The degree of discontinuity of $\textsc{1Pure}_n$ turns out to be equivalent to another family of problems, $MLPO_n$, introduced in \cite{weihrauchc} as generalizations of the \emph{lesser limited principle of omniscience} (LPO) studied in constructive mathematics (\cite{bishop}).

\begin{definition}
A function $f : \{(p_1, \ldots, p_n) \in (\mathbb{N}^\mathbb{N})^n \mid \exists i \leq n \ p_i = 0^\mathbb{N}\} \to \{1, 2, \ldots, n\}$ is in $MLPO_n$, if it fulfills $p_{f(p_1, p_2, \ldots, p_n)} = 0^\mathbb{N}$ for all valid \linebreak $(p_1, p_2, ..., p_n)$.
\end{definition}

\begin{theorem}
\label{mlpon1puren}
$MLPO_n \equiv_2 \textsc{1Pure}_n$
\begin{proof2}
First, we present a reduction from $MLPO_n$ to $\textsc{1Pure}_n$. The $n$ input bands for $MLPO_n$ can individually be translated to the $n$ relevant input bands for $\textsc{1Pure}_n$. As long as $0$ is read, the rational number $0$ will be printed. If another number is read in the $i$th step for the first time, print the rational number $- 2^{i + 1}$ from now on. All bands containing $0^\mathbb{N}$ will be translated to a $\rho$-name of 0, and all other bands to a $\rho$-name of a negative number, so a pure equilibrium corresponds to a $0$-entry.

For the other direction, all input values have to be compared. As long as no contradiction for the assumption that the $i$th value is the largest one has been found, $0$ will be printed on the $i$th output tape. If contradiction is found, print $1$. Then an output tape contains $0^\mathbb{N}$, if the corresponding input tape contains a $\rho$-name of a maximal entry.
\end{proof2}
\end{theorem}

In the next step, we extend the scope of consideration to finding pure equilibria in arbitrary bi-matrix games. The relevant problems are $\textsc{Pure}_{nm}$, where the size of the game is restricted to $n \times m$, and the general case denoted by $\textsc{Pure}$. For obtaining results, reducibility to $MLPO_n$ shall be expressed by a partition property:

\begin{lemma}
\label{criteriamlpon}
Let $H$ be a multi-valued function defined on a strongly zero-\linebreak dimensional metrisable space\footnote{Examples for such spaces are $\{0, 1\}^\mathbb{N}$ and $\mathbb{N}^\mathbb{N}$ with their standard topologies. A brief characterization of strongly zero-dimensional metrisable spaces can be found in \cite{paulyreducibilitylattice}, for details we refer to \cite{hertling} and \cite{engelking}.}  $X$. Then $H \leq_2 MLPO_n$ holds, iff there are $n$ closed sets $A_i$, $i \leq n$ with $X = \bigcup \limits_{i = 1}^n A_i$, so that for each $i \leq n$, there is an $f^i \in H$ so that $f^i_{|A_i}$ is continuous.
\begin{proof2}
Assume $H \leq_2 MLPO_n$, so there are continuous $F$, $G$ with \linebreak $x \mapsto F(x, g(G(x))) \in H$ for all $g \in MLPO_n$. Let the $i$th component of $G$ be denoted by $G_i$. Consider $G_i^{-1}(0^\mathbb{N})$. As $G_i$ is continuous, this set is closed. There is a function $\bar{g} \in MLPO_n$, so that for $x \in G_i^{-1}(0^\mathbb{N})$, $\bar{g}(G(x)) = i$ holds. Thus, if $x \mapsto F(x, \bar{g}(G(x))) \in H$ is restricted to $G_i^{-1}(0^\mathbb{N})$, it is equal to $x \mapsto F(x, i)$, and therefore it is continuous. As there is an $i$ with $G_i(x) = 0^\mathbb{N}$ for each $x \in X$, $X = \bigcup \limits_{i = 1}^n G_i^{-1}(0^\mathbb{N})$ holds, completing the first part of the proof.

For the other direction, note that for each closed subset $A$ of a strongly zero-dimensional metrisable space $X$, there is a continuous function $d_A: X \to \{0, 1\}^\mathbb{N}$ with $A = d_A^{-1}(\{0^\mathbb{N}\})$. Given sets $A_i$, $i \leq n$ as specified above, consider the function $D: X \to (\{0, 1\}^\mathbb{N})^n$ defined through $D(x)(i) = d_{A_i}(x)$. Further, define a continuous function $F$ on the set $\bigcup \limits_{i = 1}^n A_i \times \{i\}$ through $F(x, i) = f_{|A_i}^i(x)$, where $f^i \in H$ is a function that is continuous when restricted to $A_i$. As $H$ is a multi-valued function, $x \mapsto F(x, g(D(x)))$ is in $H$ for each $g \in MLPO_n$.
\end{proof2}
\end{lemma}

\begin{theorem}
\label{purenmmlponm}
$\textsc{Pure}_{nm} \leq_2  MLPO_{n*m}$.
\begin{proof2}
Given an $n \times m$ bi-matrix game $(A, B)$, the condition for the pair $(i, j)$ to be a pure equilibrium is $A_{ij} \geq A_{kj}$ and $B_{ij} \geq B_{il}$ for all $k \leq n$, $l \leq m$. This implies that the set $P_{nm}^{ij} = \{(A, B) \mid (i, j) \textnormal{ is an equilibrium of } (A, B)\} \subseteq \mathbb{R}^{nm} \times \mathbb{R}^{nm}$ is closed. Due to the admissibility of $\Gamma$, the set of corresponding names for the games is also closed. As the set of $n \times m$ bi-matrix games which have a pure strategy equilibrium is the union $\bigcup \limits_{i \leq n, j \leq m} P_{nm}^{ij}$, an application of Lemma \ref{criteriamlpon} yields the claim.
\end{proof2}
\end{theorem}

\begin{corollary}
\label{pure1pure}
$\textsc{1Pure} \equiv_2 \textsc{Pure}$.
\begin{proof2}
As both problems are the respective limits, considering Theorems \ref{mlpon1puren} and \ref{purenmmlponm} is sufficient.
\end{proof2}
\end{corollary}

The same reasoning used to establish the equivalence of finding pure strategies in 1 player games and in 2 player games can directly be extended to any finite number of players. While Nash and correlated equilibria have the same degree of discontinuity as pure equilibria in single player games, we will continue to show that a higher degree of discontinuity emerges in the two player case.

\section{Nash and correlated equilibria in bi-matrix games}
We will now consider Nash and correlated equilibria in bi-matrix games. The problems $\textsc{Corr}_{nm}$ and $\textsc{Nash}_{nm}$ are the fixed size versions, $\textsc{Corr}$ and $\textsc{Nash}$ the general problems. An additional dimension of the problem is whether the games are zero-sum, yielding the problems $\textsc{ZCorr}_{nm}$, $\textsc{ZNash}_{nm}$ and the corresponding general problems. Straight-forward reasoning yields the reductions: $$\textsc{ZCorr}_{nm} \leq_2 \textsc{Corr}_{nm} \leq_2 \textsc{Nash}_{nm} \ \ \ \textsc{ZCorr}_{nm} \leq_2 \textsc{ZNash}_{nm} \leq_2 \textsc{Nash}_{nm}$$
\subsection{The discontinuity of robust division}
Similar to $MLPO_n$ being representative of the kind of discontinuity we face when searching for pure equilibria, we will start with considering division, which will turn out to be typical for correlated and Nash equilibria. Computing $\frac{a}{b}$ given two real numbers $a$, $b \neq 0$ is continuous, of course. However, testing whether $b \neq 0$ is not. A robust variant of division, which accepts division by zero and returns an arbitrary value, is not continuous anymore:

\begin{definition}
Let $\textsc{rDiv}$ be the set of functions $d$ defined on \linebreak $\{(u, v) \mid 0 \leq \rho(u) \leq \rho(v)\}$ satisfying $\rho(d(u, v)) = \frac{\rho(u)}{\rho(v)}$ for $\rho(v) > 0$.
\end{definition}

While $\Lev(\textsc{rDiv}) = 2$ establishes robust division as an only slightly discontinuous problem, the following result shows that robust division introduces a new kind of discontinuity not present in finding pure equilibria.

\begin{theorem}
\label{divnleqnash1}
$\textsc{rDiv} \nleq_2 \textsc{Pure}$.
\begin{proof2}
We assume $\textsc{rDiv} \leq_2 \textsc{1Pure}$, due to Corollary \ref{pure1pure}. This implies the existence of continuous functions $F$, $G$, $L$ so that for each $E \in \textsc{1Pure}$ the function $d_E$ defined through $d_E(u, v) = F(u, v, E(0^{L(u, v)}101G(u, v)))$ is in $\textsc{rDiv}$. Informally, $L$ chooses the size of the game, $G$ gives the game and $F$ uses a maximal value of the game to derive the result.

We consider $n = L(0^\mathbb{N}, 0^\mathbb{N})$. As $L$ is continuous, the set $L^{-1}(\{n\})$ is open and closed, so it contains an open environment of $(0^\mathbb{N}, 0^\mathbb{N})$. Especially there is a $k \in \mathbb{N}$ with $0^k\mathbb{N}^\mathbb{N} \times 0^k\mathbb{N}^\mathbb{N} \subseteq L^{-1}(\{n\})$. We note $\frac{\rho(u)}{\rho(v)} = \frac{\rho(0^k\overline{u})}{\rho(0^k\overline{v})}$ where $\nu(\overline{u}(i)) = \nu(u(i)) * 2^{-k-1}$ and $\nu(\overline{v}(i)) = \nu(v(i)) * 2^{-k-1}$. Thus we obtain $\textsc{rDiv} \leq_2 \textsc{1Pure}_n$.

According to Lemma \ref{criteriamlpon}, $\textsc{rDiv} \leq_2 \textsc{1Pure}_n$ implies the existence of $n$ closed sets $A_i$ so that for each $i$ there is an $f_i \in \textsc{rDiv}$ so that $f_i$ restricted to $A_i$ is continuous. If there is an $l$ with $(0^\mathbb{N}, 0^\mathbb{N}) \notin A_l$, then there is a $k \in \mathbb{N}$ with $(0^k\mathbb{N}^\mathbb{N} \times 0^k\mathbb{N}^\mathbb{N}) \cap A_l = \emptyset$, so with a repetition of the argument used above we can conclude $\textsc{rDiv} \leq_2 \textsc{1Pure}_{n-1}$. Thus we can assume $(0^\mathbb{N}, 0^\mathbb{N}) \in A_l$ for all $l \leq n$.

For $l \leq n + 1$ we define a sequence $(w_k^l)_{k \in \mathbb{N}}$ of sequences through $w_k^l(i) = 0$ for $i \leq k$ and $\nu(w_k^l(i)) = (l2^k)^{-1}$ for $i > k$. Furthermore, define the sequence $(v_k)_{k \in \mathbb{N}}$ of sequences through $v_k(i) = 0$ for $i \leq k$ and $v_k(i) = 2^{-k}$ for $i > k$. For each sequence $(w_k^l, v_k)$ there must be an $l'$ so that $A_{l'}$ contains an infinite subsequence $(\overline{w}_k^l, \overline{v}_k)$ of $(w_k^l, v_k)$. As there are $n + 1$ sequences and $n$ sets, the pigeonhole principle ensures that there is a set $A_{i}$ containing the sequences $(\overline{w}_k^{l_1}, \overline{v}_k)$ and $(\overline{w}_k^{l_2}, \overline{v}_k)$.

Now observe $\lim \limits_{k \to \infty} (\overline{w}_k^{l_1}, \overline{v}_k) = \lim \limits_{k \to \infty} (\overline{w}_k^{l_2}, \overline{v}_k) = (0^\mathbb{N}, 0^\mathbb{N})$, but $f_i(\overline{w}_k^{l_1}, \overline{v}_k) = l_1^{-1} \neq l_2^{-1} = f_i(\overline{w}_k^{l_2}, \overline{v}_k)$. Thus, the restriction of $f_i$ to $A_i$ is not continuous in $(0^\mathbb{N}, 0^\mathbb{N})$, yielding a contradiction to the assumption.
\end{proof2}
\end{theorem}

We will now use modifications of the game \emph{matching pennies} as a gadget to implement divisions in a game.
$$A = \left( \begin{array}{cc} a & 0\\ 0 & b\end{array} \right) \ \ \ B = -A \ \ \ MP(a,b) = (A, B)$$
If both $a > 0$ and $b > 0$, the unique correlated equilibrium is obtained from the unique Nash equilibrium $x = y = (\frac{b}{a + b}, \frac{a}{a + b})$. If $a = 0$, $b > 0$, then $(x, y)$ is an equilibrium, iff $y = (1, 0)$, and for $a > 0$, $b = 0$ we have $y = (0, 1)$.

\begin{theorem}
\label{divleqzeco22}
$\textsc{rDiv} \leq_2 \textsc{ZCorr}_{22}$
\begin{proof2}
Given a pair of $\rho$-names for real numbers $a$, $b$ with $0 \leq a \leq b$, a name for the game $MP(a, b - a)$ can be computed. A correlated equilibrium $C$ of $MP(a, b - a)$ has the form: $$C = \left (\begin{array}{cc} c_{11} & c_{12} \\ c_{21} & c_{22}\end{array} \right ) =  \left (\begin{array}{cc} xy & x(1 - y) \\ (1 - x)y & (1 - x)(1 - y)\end{array} \right )$$ Thus, one can obtain $c_{11} + c_{21} = y = \frac{a}{b}$ for $b > 0$.
\end{proof2}
\end{theorem}

Theorem \ref{divleqzeco22} in conjunction with Theorem \ref{divnleqnash1} implies $\textsc{ZCorr}_{22} \nleq_2 \textsc{Pure}$, so even the simplest case of finding mixed strategies is not reducible to finding pure strategies. The problem $\textsc{rDiv}$ itself cannot capture the discontinuity of finding Nash equilibria, due to $\Lev(\textsc{ZNash}_{22}) = 4$ (s. Subsection \ref{levelnash22}), compelling us to derive a sequence of problems with increasing level from $\textsc{rDiv}$.

\subsection{Products of Problems and Products of Games}
The product of functions can be considered as computing all of them in parallel. This will allow us to specify exactly the degree of discontinuity of problems solvable by multiple robust divisions, once we defined products for multi-valued functions. The following definitions and results on the products of multi-valued functions and their discontinuity extend corresponding results from \cite{paulymaster}.

\begin{definition}
For functions $f: X \to Y$, $g: U \to V$, define $\langle f, g\rangle : (X \times U) \to (Y \times V)$ through $\langle f, g\rangle(x, u) = (f(x), g(u))$. Define $\langle f \rangle^1 = f$ and $\langle f \rangle^{n + 1} = \langle f, \langle f \rangle^n \rangle$.
\end{definition}

\begin{definition}
For relations $P$, $Q$, define $\langle P, Q\rangle = \{\langle f, g \rangle \mid f \in P, g \in Q\}$. Define $\langle P \rangle = P$ and $\langle P \rangle^{n + 1} = \langle P, \langle P \rangle^n \rangle$.
\end{definition}

$\lceil P, Q \rceil \leq_2 \langle P, Q\rangle$ holds, but the converse is false in general. If $f \leq_2 g$ holds, then also $\langle f, h\rangle \leq_2 \langle g, h\rangle$. As $\langle \ \rangle$ is associative, it can be extended to any finite number of arguments in the standard way. There is a useful distributive law for $\lceil \ \rceil$ and $\langle \ \rangle$ which we will state as $\langle P, \lceil Q_i \rceil_{i \in \mathbb{N}}\rangle \equiv_2 \lceil \langle P, Q_i \rangle \rceil_{i \in \mathbb{N}}$.

For games, our notion of a product will be inspired by the model of playing two independent games at once. This will allow us to establish a link between products of relations and products of games. We will use $[ \ ]$ to denote a bijection between $\{1, 2, \ldots, n\} \times \{1, 2, \ldots, m\}$ and $\{1, 2, \ldots, nm\}$ for suitable $n$, $m$.

\begin{definition}
Given an $n_1 \times m_1$ bi-matrix game $(A^1, B^1)$ and an $n_2 \times m_2$ bi-matrix game $(A^2, B^2)$, we define the $(n_1n_2) \times (m_1m_2)$ product game $(A^1, B^1) \times (A^2, B^2)$ as $(A, B)$ with $A_{[i_1, i_2][j_1, j_2]} = A_{i_1j_1}^1 + A_{i_2j_2}^2$ and \linebreak $B_{[i_1, i_2][j_1, j_2]} = B_{i_1j_1}^1 + B_{i_2j_2}^2 $.
\end{definition}

The product of games nicely commutes with the notions from game theory used in this paper, as will be established by the following theorems. A slight exception holds for the zero-sum property: A zero-sum game can always be expressed as the product of two constant-sum games which are not zero-sum. However, as a constant-sum game can always be normalized to an equivalent zero-sum game, this is not problematic for our purposes.

For simplifying notation, in the following theorems and their proofs, $(A, B)$ always abbreviates $(A^1, B^1) \times (A^2, B^2)$.
\begin{theorem}
$(A, B)$ is constant-sum, if and only if both $(A^1, B^1)$ and $(A^2, B^2)$ are constant-sum.
\begin{proof2}
Assume that $(A^1, B^1)$ and $(A^2, B^2)$ are constant-sum, that is $A^k_{ij} + B^k_{ij} = c^k$ for $k \in \{1, 2\}$ and all $i, j$. Then we have $$A_{[i_1, i_2][j_1, j_2]} + B_{[i_1, i_2][j_1, j_2]} = A_{i_1j_1}^1 + A_{i_2j_2}^2 + B_{i_1j_1}^1 + B_{i_2j_2}^2 = c^1 + c^2$$ for all $i_1, i_2, j_1, j_2$, so $(A, B)$ is also a constant-sum game.

For the other direction, we assume w.l.o.g. that $(A^1, B^1)$ is not constant-sum, so there are $i_1$, $j_1$, $k_1$, $l_1$ with $A^1_{i_1, j_1} + B^1_{i_1, j_1} \neq A^1_{k_1, l_1} + B^1_{k_1, l_1}$. Then we have: $$\begin{array}{crcl} & A_{[i_1, 1], [j_1, 1]} + B_{[i_1, 1], [j_1, 1]} & = & A^1_{i_1, j_1} + B^1_{i_1, j_1} + A^2_{1, 1} + B^2_{1, 1} \\ \neq & A^1_{k_1, l_1} + B^1_{k_1, l_1} + A^2_{1, 1} + B^2_{1, 1} & = & A_{[k_1, 1], [l_1, 1]} + B_{[k_1, 1], [l_1, 1]}\end{array}$$ Thus, the product $(A, B)$ is not constant-sum.
\end{proof2}
\end{theorem}

\begin{theorem}
If $(i_k, j_k)$ is a pure equilibrium of $(A^k, B^k)$ for $k \in \{0, 1\}$, if and only if $([i^1, i^2], [j^1, j^2])$ is a pure equilibrium of $(A, B)$.
\begin{proof2}
The proof is done by contraposition. Assume w.l.o.g. that $\hat{i}_1$ is a better response to $j_1$ than $i_1$, that is $A^1_{\hat{i}_1, j_1} > A^1_{i_1, j_1}$. Then we also have $A_{[\hat{i}_1, i_2], [j_1, j_2]} > A_{[i_1, i_2], [j_1, j_2]}$, so if $(i_1, j_1)$ is not a Nash equilibrium, then $([i^1, i^2], [j^1, j^2])$ cannot be one either.

If, on the other hand, $[\hat{i}^1, \hat{i}^2]$ is a better response against $[j^1, j^2]$ than $[i^1, i^2]$, then we have $A^1_{\hat{i}^1, j_1} + A^2_{\hat{i}_2, j_2} > A^1_{i_1, j_1} + A^2_{i_2, j_2}$. Obviously, this contradicts the conjunction of $A^1_{i_1, j_1} \geq A^1_{\hat{i}^1, j_1}$ and $A^2_{i_2, j_2} \geq A^2_{\hat{i}_2, j_2}$.
\end{proof2}
\end{theorem}

\begin{theorem}
\label{theoremnashproduct1}
If $(x^k, y^k)$ is a Nash equilibrium of $(A^k, B^k)$ for both $k \in \{0, 1\}$, then $(x, y)$ is a Nash equilibrium of $(A, B)$, where $x_{[i_1i_2]} = x_{i_1}^1x_{i_2}^2$ and $y_{[m_1m_2]} = y_{m_1}^1y_{m_2}^2$.
\begin{proof2}
We will prove that $x$ is a best response to $y$, if $x^k$ is a best response to $y^k$ for both $k \in \{0, 1\}$, the remaining part is analogous. By applying the following equivalence transformation
$$\begin{array}{cl} & \sum \limits_{o = 1}^{(n_1n_2)} \sum \limits_{p = 1}^{(m_1m_2)} x_o A_{o, p} y_p
\\ = & \sum \limits_{o_1 = 1}^{n_1} \sum \limits_{o_2 = 1}^{n_2} \sum \limits_{p_1 = 1}^{m_1} \sum \limits_{p_2 = 1}^{m_2} x_{[o_1, o_2]} A_{[o_1, o_2], [p_1, p2]} y_{[p_1, p_2]}
\\ = & \sum \limits_{o_1 = 1}^{n_1} \sum \limits_{o_2 = 1}^{n_2} \sum \limits_{p_1 = 1}^{m_1} \sum \limits_{p_2 = 1}^{m_2} x^1_{o_1}x^2_{o_2} (A^1_{o_1, p_1} + A^2_{o_2, p_2}) y^1_{p_1} y^2_{p_2}
\\= & \left [ \sum \limits_{o_1 = 1}^{n_1} \sum \limits_{p_1 = 1}^{m_1} x_{o_1} A^1_{o_1, p_1} y_{p_1} \left (\sum \limits_{o_2 = 1}^{n_2} x_{o_2} \right ) \left ( \sum \limits_{p_2 = 1}^{m_2} y_{p_2} \right ) \right ] \\ & + \left [ \sum \limits_{o_2 = 1}^{n_2} \sum \limits_{p_2 = 1}^{m_2} x_{o_2} A^2_{o_2, p_2} y_{p_2} \left (\sum \limits_{o_1 = 1}^{n_1} x_{o_1} \right ) \left ( \sum \limits_{p_1 = 1}^{m_1} y_{p_1} \right ) \right ]
\\= & \left [ \sum \limits_{o_1 = 1}^{n_1} \sum \limits_{p_1 = 1}^{m_1} x_{o_1} A^1_{o_1, p_1} y_{p_1} \right] + \left [ \sum \limits_{o_2 = 1}^{n_2} \sum \limits_{p_2 = 1}^{m_2} x_{o_2} A^2_{o_2, p_2} y_{p_2} \right ] \end{array}$$
on both sides of the best response condition $$\sum \limits_{o = 1}^{n_1n_2} \sum \limits_{p = 1}^{m_1m_2} x_o A_{o, p} y_p \geq \sum \limits_{o = 1}^{n_1n_2} \sum \limits_{p = 1}^{m_1m_2} \hat{x}_o A_{o, p} y_p$$ one obtains the form for the best response condition:
$$\begin{array}{rcl} \left [ \sum \limits_{o_1 = 1}^{n_1} \sum \limits_{p_1 = 1}^{m_1} x_{o_1} A^1_{o_1, p_1} y_{p_1} \right] & + & \left [ \sum \limits_{o_2 = 1}^{n_2} \sum \limits_{p_2 = 1}^{m_2} x_{o_2} A^2_{o_2, p_2} y_{p_2} \right ] \\ & \geq & \\ \left [ \sum \limits_{o_1 = 1}^{n_1} \sum \limits_{p_1 = 1}^{m_1} \hat{x}_{o_1} A^1_{o_1, p_1} y_{p_1} \right] & + & \left [ \sum \limits_{o_2 = 1}^{n_2} \sum \limits_{p_2 = 1}^{m_2} \hat{x}_{o_2} A^2_{o_2, p_2} y_{p_2} \right ]\end{array}$$ As this is just the sum of the best response conditions for the individual games $(A^1, B^2)$ and $(A^2, B^2)$, the claim follows.
\end{proof2}
\end{theorem}

\begin{theorem}
If $(x, y)$ is a Nash equilibrium of $(A, B)$, then $(x^1, y^1)$ given by $x_{i}^1 = \sum \limits_{l = 1}^{n_2} x_{[i,l]}$ and $y_{j}^1 = \sum \limits_{l = 1}^{m_2} y_{[j,l]}$ is a Nash equilibrium of $(A^1, B^1)$.
\begin{proof2}
Again the proof uses contraposition. Assume w.l.o.g. that $\hat{x}^1$ is a better response against $y^1$ than $x^1$, that is: $$\sum \limits_{o = 1}^{n_1} \sum \limits_{p = 1}^{m_1} \hat{x}^1_o A^1_{o,p} y^1_p >  \sum \limits_{o = 1}^{n_1} \sum \limits_{p = 1}^{m_1} x^1_o A^1_{o,p} y^1_p$$ Add $\sum \limits_{o = 1}^{n_2} \sum \limits_{p = 1}^{m_2} x^2_o A^2_{o,p} y^2_p$ on both sides, and apply the reverse of the transformation used in the proof of Theorem \ref{theoremnashproduct1}. Then one obtains: $$\sum \limits_{o = 1}^{n_1n_2} \sum \limits_{p = 1}^{m_1m_2} \hat{x}_o A_{o, p} y_p > \sum \limits_{o = 1}^{n_1n_2} \sum \limits_{p = 1}^{m_1m_2} \hat{x}_o A_{o, p} y_p$$ with $\hat{x}$ defined via $\hat{x}_{[o_1, o_2]} = \hat{x}^1_{o_1}x^2_{o_2}$. This contradicts the assumption that $x$ would be a best response against $y$, so $(x, y)$ cannot be a Nash equilibrium.
\end{proof2}
\end{theorem}

\begin{theorem}
\label{theoremcorrproduct1}
If $C^k$ is a correlated equilibrium for $(A^k, B^k)$ for both $k \in \{0, 1\}$, then a correlated equilibrium of $(A, B)$ is given by $C$, defined via $C_{[i_1, i_2], [j_1, j_2]} = C^1_{i_1, j_1}C^2_{i_2, j_2}$.
\begin{proof2}
We show the claim only for the condition for the first player, the second player's condition is dealt with in the same way. We add the inequalities $$\sum \limits_{j_k = 1}^{m_k} A^k_{i_kj_k}C^k_{i_kj_k} \geq \sum \limits_{j_k = 1}^{m_k} A^k_{l_kj_k}C^k_{i_kj_k}$$ holding for all $i_k$, $l_k$ for both $k \in \{0, 1\}$ to arrive at: $$\sum \limits_{j_1 = 1}^{m_1} A^1_{i_1j_1}C^1_{i_1j_1} + \sum \limits_{j_2 = 1}^{m_2} A^2_{i_2j_2}C^2_{i_2j_2} \geq \sum \limits_{j_1 = 1}^{m_1} A^1_{l_1j_1}C^1_{i_1j_1} + \sum \limits_{j_2 = 1}^{m_2} A^2_{l_2j_2}C^2_{i_2j_2}$$
Taking into consideration $\sum \limits_{i_k = 1}^{n_k} \sum \limits_{j_k = 1}^{m_k} C^k_{i_kj_k} = 1$ for both $k \in \{0, 1\}$, and summing over all free variables, this can be expanded to:
$$\begin{array}{cl} & \sum \limits_{i_1 = 1}^{n_1} \sum \limits_{j_1 = 1}^{m_1} \sum \limits_{i_2 = 1}^{n_2} \sum \limits_{j_2 = 1}^{m_2} \left ( A^1_{i_1j_1}C^1_{i_1j_1}C^2_{i_2j_2} + A^2_{i_2j_2}C^1_{i_1j_2}C^2_{i_2j_2} \right ) \\ \geq & \sum \limits_{i_1 = 1}^{n_1} \sum \limits_{j_1 = 1}^{m_1} \sum \limits_{i_2 = 1}^{n_2} \sum \limits_{j_2 = 1}^{m_2} \left ( A^1_{l_1j_1}C^1_{i_1j_1}C^2_{i_2j_1} + A^2_{l_2j_2}C^1_{i_1j_1}C^2_{i_2j_2} \right )\end{array}$$
Application of the definition of $C$ and $A$ yields:
$$\sum \limits_{i_1 = 1}^{n_1} \sum \limits_{j_1 = 1}^{m_1} \sum \limits_{i_2 = 1}^{n_2} \sum \limits_{j_2 = 1}^{m_2} A_{[i_1, i_2], [j_1, j_2]}C_{[i_1, i_2], [j_1, j2]} \geq \sum \limits_{i_1 = 1}^{n_1} \sum \limits_{j_1 = 1}^{m_1} \sum \limits_{i_2 = 1}^{n_2} \sum \limits_{j_2 = 1}^{m_2} A_{[l_1, l_2], [j_1, j_2]}C_{[i_1, i_2], [j_1, j2]}$$As $[ \ ]$ is bijective, this is the condition we needed to prove.
\end{proof2}
\end{theorem}

\begin{theorem}
If $C$ is a correlated equilibrium for $(A, B)$, then a correlated equilibrium of $(A^1, B^1)$ can be obtained by $C^1_{i_1, j_1} = \sum \limits_{i_2 = 1}^{n_2} \sum \limits_{j_2 = 1}^{m_2} C_{[i_1, i_2],[j_1, j_2]}$.
\begin{proof2}
The result is obtained by reversal of the proof of Theorem \ref{theoremcorrproduct1}.
\end{proof2}
\end{theorem}
As the product game can be computed from the constituent games, we can use the properties of the products of games to obtain the following results regarding the problem of finding equilibria:
\begin{theorem}
\label{productssingle}
Let $\textsc{Game} \in \{\textsc{Pure}, \textsc{ZCorr}, \textsc{ZNash}, \textsc{Corr}, \textsc{Nash}\}$. Then \linebreak $\langle \textsc{Game}_{nm}, \textsc{Game}_{kl}\rangle \leq_2 \textsc{Game}_{(nk),(ml)}$.
\end{theorem}

\begin{theorem}
\label{productssupremum}
Let $\textsc{Game} \in \{\textsc{Pure}, \textsc{ZCorr}, \textsc{ZNash}, \textsc{Corr}, \textsc{Nash}\}$. Then \linebreak $\langle \textsc{Game} \rangle^n \equiv_2 \textsc{Game}$ for all $n \in \mathbb{N}$.
\begin{proof2}
We prove the claim for $n = 2$, the remaining part is done by induction. $\textsc{Game} \leq_2 \langle \textsc{Game}, \textsc{Game} \rangle$ is trivial. Now observe $\textsc{Game} = \lceil \textsc{Game}_{nm} \rceil_{n,m \in \mathbb{N}}$ and use the distributive law twice, yielding \linebreak $\langle \textsc{Game}, \textsc{Game} \rangle \equiv_2 \lceil \langle \textsc{Game}_{nm}, \textsc{Game}_{kl}\rangle \rceil_{n,m,k,l \in \mathbb{N}}$. Application of Theorem \ref{productssingle} concludes the proof.
\end{proof2}
\end{theorem}

The present paper contains two results interpretable as counterparts to Theorem \ref{productssingle}, as they allow to reduce finding equilibria for a large game to finding equilibria in several smaller games; for mixed strategies, this will be a consequence of the main result presented in Subsection \ref{mainresult}, the corresponding statement for pure strategies is given in the next theorem:

\begin{theorem}
\label{nashleqmlpo2product}
$\textsc{1Pure}_{n+1} \leq_2 \langle MLPO_2 \rangle^{n}$.
\begin{proof2}
We describe a Type-2-Machine $M$ with $n + 1$ input bands containing the relevant payoffs, $n$ pairs of two output bands, referred to as $ia$ and $ib$ for $i \leq n$. In the $i$th step of the computation, $M$ will write $0$ on every output tape. Let $w_{ik}$ be the rational number encoded by the $i$th natural number on the $k$th input tape. Then, for each $k \leq n$, $M$ will write $1$ on the output tape $ka$, iff $w_{ik} < w_{i,k+j} + 2^{-i}$ holds for all $j \leq n + 1 - k$, and write $1$ on the output tape $kb$, iff $w_{i,k+j} < w_{i,k} + 2^{-i}$ holds for all $j \leq n + 1 - k$. Considering the definition of $\rho$, for no $i \leq n$ the symbol $1$ can be written on both $ia$ and $ib$, so the output of $M$ is in the domain of $\langle MLPO_2 \rangle^{n}$.

Assume that first $M$ and then $\langle MLPO_2 \rangle^{n}$ is applied to an $(n + 1)$-matrix game. Denote the input with $(P_1, P_2, ..., P_{n+1})$ and the output with $w$. If there is a $k \leq n$, so that $w(k) = 1$ and $w(j) = 2$ holds for $j < k$, then $\rho(P_k) \geq \rho(P_i)$ holds for all $i \leq n + 1$, so $k$ is a valid output. If $w(j) = 2$ holds for all $j \leq n$, then $n + 1$ is a valid output, as no other input value is greater than the $n + 1$st.
\end{proof2}
\end{theorem}

As we have identified $MLPO_2$ (or $\textsc{1Pure}_2$) as the basic building stone in the degree of discontinuity of finding pure strategies, the following theorem will establish the missing link in the relationship between finding pure strategies and multiple robust divisions:

\begin{theorem}
\label{mlpo2lessdiv}
$MLPO_2 <_2 \textsc{rDiv}$.
\begin{proof2}
$\textsc{rDiv} \nleq_2 MLPO_2$ has already been proven. For the other direction, note that there is a computable (and hence continuous) function turning arbitrary sequences of natural numbers into $\rho$-names, so that a sequence is mapped to a $\rho$-name of $0$ iff it is $0^\mathbb{N}$. Thus we can assume that the input of $MLPO_2$ is given as two $\rho$-names $a$, $b$ of real numbers, with at least one of them being $0$. Consider $\textsc{rDiv}(|a|, |a| + |b|)$. If this is not a $\rho$-name of $1$, then $a$ must be a $\rho$-name of $0$. If the output is not a $\rho$-name of $0$, then $b$ must be a $\rho$-name of $0$.
\end{proof2}
\end{theorem}

To sum up the results established sofar, we have: $$\lceil \langle \textsc{1Pure}_2 \rangle^n \rceil_{n \in \mathbb{N}} \equiv_2 \textsc{1Pure} \equiv_2 \textsc{Pure} <_2 \lceil \langle \textsc{rDiv} \rangle^n \rceil_{n \in \mathbb{N}} \leq_2 \textsc{ZCorr}$$

\subsection{Problems reducible to $\lceil \langle \textsc{rDiv} \rangle^n \rceil_{n \in \mathbb{N}}$}
\label{mainresult}
The goal of this subsection is to present a way of designing reductions to \linebreak $\lceil \langle \textsc{rDiv} \rangle^n \rceil_{n \in \mathbb{N}}$, and, in particular, to present a reduction from $\textsc{Nash}$. This equivalently can be considered as the task to design algorithms for a Type-2-Machine capable of making a finite number of independent queries to an oracle for $\textsc{rDiv}$. Due to Theorems \ref{nashleqmlpo2product}, \ref{mlpo2lessdiv} also oracle calls to $MLPO_n$ are permitted.

We will start by providing a technical lemma similar to Lemma \ref{criteriamlpon}. Using the lemma, we can prove that the Fourier-Motzkin-algorithm (\cite{kessler}) for solving systems of linear inequalities can be executed using continuous (even computable) operations and oracle calls to \textsc{rDiv}.
\begin{lemma}
\label{criteriardivn}
Let $F$ be a multi-valued function defined on a strongly zero-\linebreak dimensional metrisable space $X$. Then $F \leq_2 \lceil \langle \textsc{rDiv} \rangle^n \rceil_{n \in \mathbb{N}}$ holds, iff there are $k$ closed sets $A_i$, $i \leq k$ with $X = \bigcup \limits_{i = 1}^k A_i$, so that for each $i \leq k$, there is a multi-valued function $G^i \leq_2 \lceil \langle \textsc{rDiv} \rangle^n \rceil_{n \in \mathbb{N}}$ with $\dom(G^i) = X$, so that for each $g^i \in G^i$ there is an $f^i \in F$ with $f^i_{|A_i} = g^i_{|A_i}$.
\begin{proof2}
One direction of the proof is trivial. For the other direction, assume that there are continuous functions $R_i$, $Q_i$ for each $i \leq k$ with $x \mapsto R_i(x, q(Q_i(x))) \in G^i$ for each $q \in \textsc{rDiv}$. Further let $D$ be the continuous function defined for the sets $A_i$ as in the proof to Lemma \ref{criteriamlpon}. Define the function $Q = (D, Q_1, \ldots, Q_k)$ and $R(x, i, y_1, \ldots, y_k) = R_i(x, y_i)$. Both $Q$ and $R$ are continuous, and satisfy $x \mapsto R(x, \hat{q}(Q(x))) \in F$ for $\hat{q} \in \langle MLPO_k, \langle \lceil \langle \textsc{rDiv} \rangle^n \rceil_{n \in \mathbb{N}} \rangle^k \rangle$, so we have \linebreak $F \leq_2 \langle MLPO_k, \langle \lceil \langle \textsc{rDiv} \rangle^n \rceil_{n \in \mathbb{N}} \rangle^k \rangle \equiv_2 \lceil \langle \textsc{rDiv} \rangle^n \rceil_{n \in \mathbb{N}}$.
\end{proof2}
\end{lemma}

\begin{definition}
The problem $\textsc{BLinIneq}_{nm}$ asks for a $\rho^m$-name of a vector $v$ of reals, so that $Av \leq b$ holds in addition to $0 \leq v \leq 1$, given a $\rho^{nm}$-name for a matrix $A$ and a $\rho^n$-name for a vector $b$, provided that a solution exists. For simplicity, we assume that $Av \leq b$ always contains $0 \leq v \leq 1$. $\textsc{BLinIneq}$ is the problem without fixed values $n$, $m$.
\end{definition}

\begin{theorem}
\label{linearinequalities}
$\textsc{BLinIneq} \leq_2 \lceil \langle \textsc{rDiv} \rangle^z \rceil_{z \in \mathbb{N}}$.
\begin{proof2}
As $\textsc{BLinIneq}$ is expressible as a supremum, it suffices to prove \linebreak $\textsc{BLinIneq}_{nm} \leq_2 \lceil \langle \textsc{rDiv} \rangle^z \rceil_{z \in \mathbb{N}}$ for all $n, m \in \mathbb{N}$. For this, we use induction over $m$. The case $m = 0$ is trivial, so we assume $\textsc{BLinIneq}_{n(m-1)} \leq_2 \lceil \langle \textsc{rDiv} \rangle^z \rceil_{z \in \mathbb{N}}$.

For each $K \subseteq \{1, \ldots, n\}$, abbreviate $K^C := \{1, \ldots, n\} \setminus K$. The set $D_K = \{(A, b) \mid \forall k \in K \ a_{k1} \geq 0 \wedge \forall l \in K^C \ a_{l1} \leq 0\}$ is closed, and the union $\bigcup \limits_{K \subseteq \{1, \ldots, n\}} D_K$ covers the domain of $\textsc{BLinIneq}_{nm}$. So due to Lemma \ref{criteriardivn}, it is sufficient to show that $\textsc{BLinIneq}_{nm}$ restricted to $D_K$ is reducible to $\lceil \langle \textsc{rDiv} \rangle^z \rceil_{z \in \mathbb{N}}$ for arbitrary $K \subseteq \{1, \ldots, n\}$. In the next step we assume $K$ to be fixed. With the same argument we can assume $|a_{k1}| \geq |a_{(k+1)1}|$ by renumbering the inequalities for each fixed sequence of increasing first coefficients.

Now we rewrite the given inequalities as $a_{k1}v_1 \leq b_k - \sum \limits_{i = 2}^m a_{ki}v_i$ for $k \in K$ and $-b_j + \sum \limits_{i = 2}^m a_{ji}v_i \leq -a_{j1}v_1$ for $j \in K^C$. For each pair $k \in K$, $j \in K^C$, the corresponding inequalities can be multiplied by $-a_{j1}$ respective $a_{k1}$, and then contracted to: $$a_{k1}(-b_j + \sum \limits_{i = 2}^m a_{ji}v_i) \leq -a_{j1}(b_k - \sum \limits_{i = 2}^m a_{ki}v_i)$$ Every solution to the newly created system of linear inequalities can be extended to a solution to the original system by choosing a suitable value for $v_1$. Due to the induction assumption, such a solution can be obtained by making oracle calls to $\lceil \langle \textsc{rDiv} \rangle^z \rceil_{z \in \mathbb{N}}$.

If all $a_{k1}$ were known to be non-zero, after solutions for the $v_i$, $i \geq 2$ have been obtained, $v_1$ is chosen as a solution to: $$\max \limits_{k \in K} \frac{-b_j + \sum \limits_{i = 2}^m a_{ji}v_i}{-a_{j1}} \leq v_1 \leq \min \limits_{j \in K^C} \frac{b_k - \sum \limits_{i = 2}^m a_{ki}v_i}{a_{k1}}$$
If a certain coefficient $a_{k1}$ is $0$, then the corresponding inequality does not restrict the value of $v_1$ at all, thus, divisions by $0$ should be ignored in the expression above. However, since we cannot determine whether a certain coefficient is $0$ or not, this approach cannot be employed by us.

Therefore, we evaluate all expressions determining $v_1$ in the order of decreasing absolute values of the coefficients, that is the expression $a_{k1}$ is considered before $a_{(k+1),1}$. This ensures that all relevant inequalities are met. Thus, to obtain a solution for $v_1$, we would like to call $$v_1 = \max(0, \min(1, \op_1(\textsc{rDiv}(|b_1 - \sum \limits_{i = 2}^m a_{1i}v_i|, |a_{11}|), \op_2(|\textsc{rDiv}(b_2 - \sum \limits_{i = 2}^m a_{2i}v_i|, |a_{21}|), \ldots$$ with $\op_i = \min$ for $i \in K$ and $\op_i = \max$ else. As the $|a_{k1}|$ are ordered as a decreasing sequence, values that arise arbitrary as result of a division by zero occur deeper inside the nested structure than significant values. While they can influence the actual value for $v_1$ that is chosen, it still satisfies all inequalities, if this is possible. However, the expression above contains nested calls to $\textsc{rDiv}$ in form of the $v_i$, $2 \leq i \leq n$.

To solve the problem, one replaces $v_2$ with the corresponding sequence used to compute it, then $v_3$, and so on. By moving the $\max$ and $\min$ operators outside, and unifying all divisions, terms of the form $\textsc{rDiv}(P, Q)$ remain, where $P$ is a polynomial in $a_{ij}$, $b_j$ whose degree does not exceed $2n$, and $Q$ is a polynomial in $a_{ij}$ whose degree does not exceed $n$. These can be evaluated by allowed oracle calls, and the $\max$ and $\min$ operators are continuous.
\end{proof2}
\end{theorem}

As the problem $\textsc{BLinIneq}$ is of considerable interest on its own, we shall note that the converse statement to Theorem \ref{linearinequalities} is also true:
\begin{theorem}
\label{linearinequalities2}
$\lceil \langle \textsc{rDiv} \rangle^z \rceil_{z \in \mathbb{N}} \leq_2 \textsc{BLinIneq}$.
\begin{proof2}
We have to show $\langle \textsc{rDiv} \rangle^n \leq_2 \textsc{BLinIneq}$ for each $n \in \mathbb{N}$. Given $n$ pairs of reals $(p_i, q_i)$, consider the system of linear equalities given by $A_{ii} = q_i$, $A_{ij} = 0$ for $i \neq j$ and $b_i = p_i$. The only solution is given by $v_i = \frac{p_i}{q_i}$. By replacing every equality with two inequalities, the needed reduction is found.
\end{proof2}
\end{theorem}

By adapting \cite[Algorithm 3.4]{agtc} and applying Lemma \ref{criteriardivn} and Theorem \ref{linearinequalities} we proceed to prove the main theorem of this subsection. Again, the reasoning directly extends to more than two players.

\begin{theorem}
\label{maintheorem}
$\textsc{Nash} \leq_2 \lceil \langle \textsc{rDiv} \rangle^z \rceil_{z \in \mathbb{N}}$.
\begin{proof2}
By the same reasoning as above, since $\textsc{Nash}$ is the supremum \linebreak $\lceil \textsc{Nash}_{nm} \rceil_{n, m \in \mathbb{N}}$, it suffices to show $\textsc{Nash}_{nm} \leq_2 \lceil \langle \textsc{rDiv} \rangle^z \rceil_{z \in \mathbb{N}}$ for arbitrary $n, m \in \mathbb{N}$.

By the best response condition (\cite[Proposition 3.1]{agtc}), a pair of mixed strategies $(x, y)$ is a Nash equilibrium of a game if each pure strategy played with positive probability in $x$ (in $y$) is a best response against $y$ (against $x$). This condition can be formalized by noting that the following set is the set of games and their Nash equilibria with support in $I$, $J$: $$\hat{G}_{I, J} = \begin{array}{r}\{(A, B, x, y) \mid j, k \in J \ l \notin J \ (x^TB)_j = (x^TB)_k \geq (x^TB)_l \ y_l = 0 \ i, p \in I \\ q \notin I \ (Ay)_i = (Ay)_p \geq (Ay)_q \ x_q = 0\}\end{array}$$
The set $\hat{G}_{I, J}$ is closed, and so is its projection $G_{I, J} = \{(A, B) \mid \exists x, y \ (A, B, x, y) \in \hat{G}_{I, J}\}$.

As every game has a Nash equilibrium, the sets $G_{I, J}$ cover the domain of $\textsc{Nash}$, so we can apply Lemma \ref{criteriardivn}. To recover the Nash equilibrium $(x, y)$ from $I$, $J$ the corresponding system of linear inequalities has to be solved, which is reducible to $\lceil \langle \textsc{rDiv} \rangle^z \rceil_{z \in \mathbb{N}}$ as established in Theorem \ref{linearinequalities}.
\end{proof2}
\end{theorem}

\begin{corollary}
$\textsc{ZCorr} \equiv_2 \textsc{Corr} \equiv_2 \textsc{ZNash} \equiv_2 \textsc{Nash} \equiv_2 \lceil \langle \textsc{rDiv} \rangle^n \rceil_{n \in \mathbb{N}}$.
\end{corollary}

The same technique applied in the proof of Theorem \ref{linearinequalities} can also be used to show that Gaussian Elimination can be reduced to $\lceil \langle \textsc{rDiv} \rangle^n \rceil_{n \in \mathbb{N}}$. This shows that the reduction of Gaussian Elimination to the rank of a matrix given in \cite{ziegler} is strict, taking into consideration Corollary \ref{fnleqnash}.

\section{Additional Results}
\subsection{$\textsc{Nash}$ and $Sep$}
To shed further light on the degree of discontinuity of $\textsc{Nash}$, we will compare it to the problem $Sep$ studied in \cite{gherardi}.
\begin{definition}
$f \in Sep$ holds, iff $f$ is a function from $$\{(p, q) \in \mathbb{N}^\mathbb{N} \times \mathbb{N}^\mathbb{N} \mid \forall n, m \in \mathbb{N} \ p(n) \neq q(m)\}$$ to $\mathbb{N}^\mathbb{N}$ satisfying $f(p(n)) = 0$ and $f(q(n)) = 1$ for all $n \in \mathbb{N}$.
\end{definition}
The problem $Sep$ was shown to be equivalent to finding an infinite path in an infinite binary tree and extending a linear functional from a subspace of a Banach space to the complete space following the Hahn-Banach Theorem. $Sep$ can be reduced to $\{C_1\}$, which is defined through $C_1(p)(n) = 1$, iff there is an $i \in \mathbb{N}$ with $p(i) = n$ and $C_1(p)(n) = 0$ else. The function $C_1$ has been introduced in \cite{stein}. In \cite[Theorem 5.5]{brattka}, it was proven that a function is $\sum_2^0$-measurable, iff it is reducible to $C_1$.

In \cite{gherardi}, $\{cf\} \nleq_2 Sep$ was shown, which can directly to extended to prove $\{f\} \nleq_2 Sep$ for all discontinuous functions $f$. In the following, we will prove that $Nash$ is strictly reducible to $Sep$, thereby obtaining a lower bound for $Sep$. For this aim, we need the level of $Sep$.

\begin{theorem}
\label{levelsepnotexistent}
$\Lev^2(Sep)$ does not exist.
\begin{proof2}
We have to prove that for every $f \in Sep$, $\Lev^2(f)$ does not exist. For notation, we define $R_p = \{p(n) \mid n \in \mathbb{N}\}$. We study the points $(p, q)$ where $f$ might be continuous. First we assume there exists an $n \in \mathbb{N} \setminus (R_p \cup R_q)$. Define $p_k$ through $p_k(i) = p(i)$ for $i \neq k$ and $p_k(k) = n$, and $q_k$ through $q_k(i) = q(i)$ for $i \neq k$ and $q_k(k) = n$. Then $\lim \limits_{k \to \infty} (p_k, q) = \lim \limits_{k \to \infty} (p, q_k) = (p, q)$ holds. However, $f(p_k, q)(n) = 0$ and $f(p, q_k)(n) = 1$, so $f$ cannot be continuous in $(p, q)$. Whether $f$ is continuous in a point $(p, q)$ with $R_p \cup R_q = \mathbb{N}$ depends on $f$.

Rephrasing the considerations above, we know $cl \{(p, q) \mid R_q \cup R_p \neq \mathbb{N}, R_q \cap R_p = \emptyset\} \subseteq \mathcal{L}^2_1(f)$. However, for each $(p, q)$ with $R_p \cup R_q = \mathbb{N}$, define $p_k$ through $p_k(i) = p(i)$ for $i \leq k$, and $p_k(i) = p(k)$ for $i \geq k$, analogously define $q_k$ through $q_k(i) = q(i)$ for $i \leq k$, and $q_k(i) = q(k)$ for $i \geq k$. Then $\lim \limits_{k \to \infty} (p_k, q_k) = (p, q)$, and for each $k$, $R_{p_k} \cup R_{q_k}$ is even finite. Thus, $\mathcal{L}_1(f) = \mathcal{L}_0(f)$ follows. Transfinite induction easily proves $\mathcal{L}^2_\alpha(f) = \mathcal{L}^2_0(f) \neq \emptyset$ for all ordinal numbers $\alpha$.
\end{proof2}
\end{theorem}

Due to the behaviour of the level under formation of products (\cite{paulymaster}) and suprema (\cite{paulyreducibilitylattice}, \cite{hertling}), we know $\Lev^2(\textsc{Nash}) = \omega$, where $\omega$ is the smallest infinite ordinal. This is sufficient to establish $Sep \nleq_2 \textsc{Nash}$ by \cite[Theorem 5.7]{paulyreducibilitylattice}.

\begin{theorem}
\label{rdivleqsep}
$\textsc{rDiv} \leq_2 Sep$.
\begin{proof2}[To Theorem \ref{rdivleqsep}]
Let $(a, b)$ be the input for $\textsc{rDiv}$. We describe a machine transforming it to an input for $Sep$. The machine has two different modes, starting in the first one. In each stage $i$, write $1$ on the first and $2$ on the second output tape. Then test $\nu(b_i) > 2^{-i + 1}$. If \textsc{yes}, change to mode 2, otherwise continue with stage $i + 1$.

In the following description of the second mode, let $i_0$ denote the stage of the first mode during which the switch occurred. We will use the following inequality: $$\left | \frac{\nu(a_i)}{\nu(b_i)} - \frac{\rho(a)}{\rho(b)} \right | \leq 2^{-\left [i - 2(i_0 + 1)\right ]}$$
The stages in the second mode are indexed by $[i, j]$ (employing a monotone bijective pairing $[ \ ]: \mathbb{N} \times \mathbb{N} \to \mathbb{N}$). In each stage, test $|\frac{\nu(a_k)}{\nu(b_k)} - \nu(j)| < 2^{-i - 1}$ with $k = i + 2i_0 + 3$. If the answer is \textsc{yes}, write $2 + [i, j]$ on the second output tape, otherwise write $2 + [i, j]$ on the first tape. Note that \textsc{yes} also implies $|\frac{\rho(a)}{\rho(b)} - \nu(j)| < 2^i$.

It remains to describe a second Type-2-Machine which recovers a $\rho$-name for $\frac{\rho{a}}{\rho{b}}$ from the result $w$ of the applying of $Sep$ to the output of the machine described above. For that, we fix a function $\kappa: \mathbb{N} \to \mathbb{N}$ with the following property: $$\forall i \ \forall k \ \exists j \leq \kappa(i) \ |\nu(k) - \nu(j)| < 2^{-i - 1}$$ If the numbering $\nu$ is chosen in a way that no computable $\kappa$ exists, $\kappa$ is assumed to be given by an oracle, this still allows a continuous reduction.

In the stage $i$ of our second machine, test $w_{2 + [i,j]} = 1?$ for each $j$ from $1$ to $\kappa(i)$. If a $j$ is found which yields $\textsc{yes}$, then write $j$ on the output tape and proceed with the next stage. If the answer is always \textsc{no}, repeat the last output forever (always write 1 if there has been no output yet).

If $\rho(b) \neq 0$, then the first machine will eventually switch to the second mode. There, all $2^{-i}$-approximations $\nu(j)$ for $\frac{\rho(a)}{\rho(b)}$ are computed. Due to the properties of $\kappa$, at least one of them fulfills $j \leq \kappa(i)$, and is found and printed by the second machine. If $\rho(b) = 0$, then the second machine is still guaranteed to produce a valid $\rho$-name.

\end{proof2}
\end{theorem}

\begin{theorem}
\label{sepsepleqsep}
$\langle Sep, Sep \rangle \equiv_2 Sep$.
\begin{proof2}
One direction is trivial. For the other, define the continuous function $G: \mathbb{N}^\mathbb{N} \times \mathbb{N}^\mathbb{N} \to \mathbb{N}^\mathbb{N}$ via $G(p, q)(2k) = 2p(k)$ and $G(p, q)(2k + 1) = 2q(k) + 1$. Define the continuous function $F: \mathbb{N}^\mathbb{N} \to \mathbb{N}^\mathbb{N} \times \mathbb{N}^\mathbb{N}$ through $F(p) = (q_1, q_2)$, where $q_1(i) = p(2i)$ and $q_2(i) = p(2i+1)$. Now observe \linebreak $\langle Sep, Sep \rangle(p_1, q_1, p_2, q_2) = F(Sep(G(p_1, p_2), G(q_1, q_2)))$.
\end{proof2}
\end{theorem}

\begin{corollary}
\label{nashsmallersep}
$\textsc{Nash} <_2 Sep$.
\begin{proof2}
Due to Theorem \ref{maintheorem} and the properties of suprema, we just have to show $\langle \textsc{rDiv} \rangle^n \leq_2 Sep$ for all $n \in \mathbb{N}$. Theorem \ref{rdivleqsep} yields $\langle \textsc{rDiv} \rangle^n \leq_2 \langle Sep \rangle^n$, repeated application of Theorem \ref{sepsepleqsep} yields $\langle Sep \rangle^n \equiv_2 Sep$.
\end{proof2}
\end{corollary}

\begin{corollary}
\label{fnleqnash}
$\{f\} \nleq_2 \textsc{Nash}$ for all discontinuous functions $f$.
\end{corollary}

\subsection{The Level of $\textsc{Nash}_{22}$}
\label{levelnash22}
The simplest non-trivial bi-matrix games, $2 \times 2$ games, have already been investigated from a constructive point of view in \cite{bridges}. Among other results, \cite{bridges} contains the constructive analogue to the reduction $MLPO_2 \leq_2 \textsc{Nash}_{22}$, and the constructive analogue to determine a subset of $\mathcal{L}_0(\textsc{Nash}_{22}) \setminus \mathcal{L}_1(\textsc{Nash}_{22})$, that is the set where Nash equilibria are continuous. We will produce the TTE-counterpart by investigating the Level of $\textsc{Nash}_{22}$.

\begin{theorem}
\label{levnash22}
$\Lev(\textsc{Nash}_{22}) = 4$.
\begin{proof2}[To Theorem \ref{levnash22}]
A strategy in this setting can be describe by merely a single real number, the probability of the first pure strategy to be chosen. This allows to condense the information given in the game: Row-player tries to maximize $x(c + dy)$ and column-player tries to maximize $y(e + fy)$ with $c = a_{11} - a_{12} - a_{21} + a_{22}$, $d = a_{12} - a_{22}$, $e = b_{11} - b_{12} - b_{21} + b_{22}$ and $f = b_{21} - b_{22}$. The table in the Appendix defines a function in $\textsc{Nash}_{22}$ depending on these value which obviously has Level $4$, establishing $\Lev(\textsc{Nash}_{22}) \leq 4$. It is straight-forward to check that all points of discontinuity of this function are points of discontinuity of every choice function for $\textsc{Nash}_{22}$.
\end{proof2}
\end{theorem}

\appendix
\section{Appendix}
$$\begin{array}{cccc|cc}
c > 0   &   d + c > 0   &   e > 0   &   e + f > 0   &   1   &   1\\
c > 0   &   d + c > 0   &   e > 0   &   e + f < 0   &   1   &   0\\
c > 0   &   d + c > 0   &   e < 0   &   e + f > 0   &   1   &   1\\
c > 0   &   d + c > 0   &   e < 0   &   e + f < 0   &   1   &   0\\
c > 0   &   d + c < 0   &   e > 0   &   e + f > 0   &   0   &   1\\
c > 0   &   d + c < 0   &   e > 0   &   e + f < 0   &   0   &   1\\
c > 0   &   d + c < 0   &   e < 0   &   e + f > 0   &   \frac{-e}{f}   &   \frac{-c}{d}\\
c > 0   &   d + c < 0   &   e < 0   &   e + f < 0   &   1   &   0\\
c < 0   &   d + c > 0   &   e > 0   &   e + f > 0   &   1   &   1\\
c < 0   &   d + c > 0   &   e > 0   &   e + f < 0   &   \frac{-e}{f}   &   \frac{-c}{d}\\
c < 0   &   d + c > 0   &   e < 0   &   e + f > 0   &   0   &   0\\
c < 0   &   d + c > 0   &   e < 0   &   e + f < 0   &   0   &   0\\
c < 0   &   d + c < 0   &   e > 0   &   e + f > 0   &   0   &   1\\
c < 0   &   d + c < 0   &   e > 0   &   e + f < 0   &   0   &   1\\
c < 0   &   d + c < 0   &   e < 0   &   e + f > 0   &   0   &   0\\
c < 0   &   d + c < 0   &   e < 0   &   e + f < 0   &   0   &   0\\
\hline
c = 0   &   d + c > 0   &   e > 0   &   e + f > 0   &   1   &   1\\
c = 0   &   d + c > 0   &   e > 0   &   e + f < 0   &   1   &   0\\
c = 0   &   d + c > 0   &   e < 0   &   e + f > 0   &   1   &   1\\
c = 0   &   d + c > 0   &   e < 0   &   e + f < 0   &   1   &   0\\
c = 0   &   d + c < 0   &   e > 0   &   e + f > 0   &   0   &   1\\
c = 0   &   d + c < 0   &   e > 0   &   e + f < 0   &   0   &   1\\
c = 0   &   d + c < 0   &   e < 0   &   e + f > 0   &   0   &   0\\
c = 0   &   d + c < 0   &   e < 0   &   e + f < 0   &   1   &   0\\
c > 0   &   d + c = 0   &   e > 0   &   e + f > 0   &   1   &   1\\
c > 0   &   d + c = 0   &   e > 0   &   e + f < 0   &   1   &   0\\
c > 0   &   d + c = 0   &   e < 0   &   e + f > 0   &   1   &   1\\
c > 0   &   d + c = 0   &   e < 0   &   e + f < 0   &   1   &   0\\
c < 0   &   d + c = 0   &   e > 0   &   e + f > 0   &   0   &   1\\
c < 0   &   d + c = 0   &   e > 0   &   e + f < 0   &   0   &   1\\
c < 0   &   d + c = 0   &   e < 0   &   e + f > 0   &   0   &   0\\
c < 0   &   d + c = 0   &   e < 0   &   e + f < 0   &   0   &   0\\
c > 0   &   d + c > 0   &   e = 0   &   e + f > 0   &   1   &   1\\
c > 0   &   d + c > 0   &   e = 0   &   e + f < 0   &   1   &   0\\
c > 0   &   d + c < 0   &   e = 0   &   e + f > 0   &   0   &   1\\
c > 0   &   d + c < 0   &   e = 0   &   e + f < 0   &   1   &   0\\
c < 0   &   d + c > 0   &   e = 0   &   e + f > 0   &   1   &   1\\
c < 0   &   d + c > 0   &   e = 0   &   e + f < 0   &   0   &   0\\
c < 0   &   d + c < 0   &   e = 0   &   e + f > 0   &   0   &   1\\
c < 0   &   d + c < 0   &   e = 0   &   e + f < 0   &   0   &   0\\
c > 0   &   d + c > 0   &   e > 0   &   e + f = 0   &   1   &   1\\
c > 0   &   d + c > 0   &   e < 0   &   e + f = 0   &   1   &   0\\
c > 0   &   d + c < 0   &   e > 0   &   e + f = 0   &   0   &   1\\
c > 0   &   d + c < 0   &   e < 0   &   e + f = 0   &   1   &   0\\
c < 0   &   d + c > 0   &   e > 0   &   e + f = 0   &   1   &   1\\
c < 0   &   d + c > 0   &   e < 0   &   e + f = 0   &   0   &   0\\
c < 0   &   d + c < 0   &   e > 0   &   e + f = 0   &   0   &   1\\
c < 0   &   d + c < 0   &   e < 0   &   e + f = 0   &   0   &   0
\end{array} \ \ \begin{array}{cccc|cc}
c = 0   &   d + c = 0   &   e > 0   &   e + f > 0   &   1   &   1\\
c = 0   &   d + c = 0   &   e > 0   &   e + f < 0   &   1   &   0\\
c = 0   &   d + c = 0   &   e < 0   &   e + f > 0   &   1   &   1\\
c = 0   &   d + c = 0   &   e < 0   &   e + f < 0   &   1   &   0\\
c = 0   &   d + c > 0   &   e = 0   &   e + f > 0   &   1   &   1\\
c = 0   &   d + c > 0   &   e = 0   &   e + f < 0   &   1   &   0\\
c = 0   &   d + c < 0   &   e = 0   &   e + f > 0   &   0   &   0\\
c = 0   &   d + c < 0   &   e = 0   &   e + f < 0   &   0   &   0\\
c = 0   &   d + c > 0   &   e > 0   &   e + f = 0   &   1   &   1\\
c = 0   &   d + c > 0   &   e < 0   &   e + f = 0   &   1   &   1\\
c = 0   &   d + c < 0   &   e > 0   &   e + f = 0   &   0   &   1\\
c = 0   &   d + c < 0   &   e < 0   &   e + f = 0   &   0   &   0\\
c < 0   &   d + c = 0   &   e = 0   &   e + f > 0   &   0   &   1\\
c < 0   &   d + c = 0   &   e = 0   &   e + f < 0   &   0   &   0\\
c > 0   &   d + c = 0   &   e = 0   &   e + f > 0   &   1   &   1\\
c > 0   &   d + c = 0   &   e = 0   &   e + f < 0   &   1   &   0\\
c < 0   &   d + c = 0   &   e > 0   &   e + f = 0   &   0   &   1\\
c < 0   &   d + c = 0   &   e < 0   &   e + f = 0   &   0   &   0\\
c > 0   &   d + c = 0   &   e > 0   &   e + f = 0   &   1   &   1\\
c > 0   &   d + c = 0   &   e < 0   &   e + f = 0   &   1   &   0\\
c > 0   &   d + c > 0   &   e = 0   &   e + f = 0   &   1   &   1\\
c > 0   &   d + c < 0   &   e = 0   &   e + f = 0   &   0   &   1\\
c < 0   &   d + c > 0   &   e = 0   &   e + f = 0   &   1   &   1\\
c < 0   &   d + c < 0   &   e = 0   &   e + f = 0   &   0   &   1\\
\hline
c = 0   &   d + c = 0   &   e = 0   &   e + f > 0   &   1   &   1\\
c = 0   &   d + c = 0   &   e = 0   &   e + f < 0   &   1   &   0\\
c = 0   &   d + c = 0   &   e < 0   &   e + f = 0   &   1   &   0\\
c = 0   &   d + c = 0   &   e > 0   &   e + f = 0   &   1   &   1\\
c = 0   &   d + c > 0   &   e = 0   &   e + f = 0   &   1   &   1\\
c = 0   &   d + c < 0   &   e = 0   &   e + f = 0   &   0   &   1\\
c < 0   &   d + c = 0   &   e = 0   &   e + f = 0   &   0   &   1\\
c > 0   &   d + c = 0   &   e = 0   &   e + f = 0   &   1   &   1\\
\hline
c = 0   &   d + c = 0   &   e = 0   &   e + f = 0   &   1   &   0
\end{array}$$

\end{document}